\documentclass[reprint, aps, amsmath, amssymb, prl, superscriptaddress]{revtex4-2}
\usepackage{graphicx}
\usepackage{amsmath}
\usepackage{hyperref}
\hypersetup{colorlinks=true, citecolor=blue}

\usepackage{xcolor}

\begin{document}
\title{Suppression of core turbulence by profile shaping in Wendelstein 7-X}

\author{A. v. Stechow}
\affiliation{Max-Planck Institute for Plasma Physics, 17491 Greifswald, Germany}


\author{O. Grulke}
\affiliation{Max-Planck Institute for Plasma Physics, 17491 Greifswald, Germany}
\affiliation{Tech. Univ. Denmark, Dept Phys, DK-2800 Lyngby, Denmark}

\author{Th. Wegner}
\affiliation{Max-Planck Institute for Plasma Physics, 17491 Greifswald, Germany}

\author{J. H. E. Proll}
\affiliation{Science and Technology of Nuclear Fusion, Department of Applied Physics,
Eindhoven University of Technology, The Netherlands}

\author{J. A. Alcus\'on}
\affiliation{Max-Planck Institute for Plasma Physics, 17491 Greifswald, Germany}

\author{H. M. Smith}
\affiliation{Max-Planck Institute for Plasma Physics, 17491 Greifswald, Germany}


\author{J. Baldzuhn}
\affiliation{Max-Planck Institute for Plasma Physics, 17491 Greifswald, Germany}

\author{C. D. Beidler}
\affiliation{Max-Planck Institute for Plasma Physics, 17491 Greifswald, Germany}

\author{M. N. A. Beurskens}
\affiliation{Max-Planck Institute for Plasma Physics, 17491 Greifswald, Germany}

\author{S. A. Bozhenkov}
\affiliation{Max-Planck Institute for Plasma Physics, 17491 Greifswald, Germany}

\author{E. Edlund}
\affiliation{SUNY Cortland, Cortland, USA}

\author{B. Geiger}
\affiliation{University of Wisconsin, Madison, USA}

\author{Z. Huang}
\affiliation{MIT, Plasma Sci. \& Fus. Ctr, 77 Massachusetts Ave, Cambridge, MA 
02139 USA}

\author{O. P. Ford}
\affiliation{Max-Planck Institute for Plasma Physics, 17491 Greifswald, Germany}

\author{G. Fuchert}
\affiliation{Max-Planck Institute for Plasma Physics, 17491 Greifswald, Germany}

\author{A. Langenberg}
\affiliation{Max-Planck Institute for Plasma Physics, 17491 Greifswald, Germany}

\author{N. Pablant}
\affiliation{Princeton Plasma Physics Laboratory, Princeton, NJ, USA}

\author{E. Pasch}
\affiliation{Max-Planck Institute for Plasma Physics, 17491 Greifswald, Germany}

\author{M. Porkolab}
\affiliation{MIT, Plasma Sci. \& Fus. Ctr, 77 Massachusetts Ave, Cambridge, MA 
02139 USA}

\author{K. Rahbarnia}
\affiliation{Max-Planck Institute for Plasma Physics, 17491 Greifswald, Germany}

\author{J. Schilling}
\affiliation{Max-Planck Institute for Plasma Physics, 17491 Greifswald, Germany}

\author{E. R. Scott}
\affiliation{University of Wisconsin, Madison, USA}

\author{H. Thomsen}
\affiliation{Max-Planck Institute for Plasma Physics, 17491 Greifswald, Germany}

\author{L. Van\'o}
\affiliation{Max-Planck Institute for Plasma Physics, 17491 Greifswald, Germany}

\author{G. Weir}
\affiliation{Max-Planck Institute for Plasma Physics, 17491 Greifswald, Germany}

\author{The W7-X Team}
\affiliation{Max-Planck Institute for Plasma Physics, 17491 Greifswald, Germany}

\begin{abstract}
In the Wendelstein 7-X magnetic confinement experiment, a reduction of turbulent density fluctuations as well as anomalous impurity diffusion is associated with a peaking of the plasma density profile.
These effects correlate with improved confinement and appear largely due to a reduction of anomalous transport as the change in neoclassical transport is small.
The observed decrease of turbulent heat flux with increased density gradients is in agreement with nonlinear gyrokinetic simulations, and has been attributed to the unique geometry of W7-X that limits the severity of trapped electron modes.
\end{abstract}

\maketitle

\paragraph{Introduction}\label{intro}
The development of magnetic confinement concepts towards a fusion reactor crucially relies on a comprehensive understanding and control of particle and energy transport processes.
Transport based on binary collisions and drifts is generally not sufficient to describe the large transport coefficients inferred from experiments.
Particularly in tokamaks, radial transport based on microturbulence driven by gradients of the kinetic plasma profiles, i.e. electron and ion temperature as well as plasma density, dominates radial plasma losses during steady-state operation and limits the central plasma parameters (for a review see \cite{Doyle07}).
In low-beta plasmas, three electrostatic instabilities usually govern turbulence, namely the electron and ion temperature gradient (ETG/ITG) and density-gradient-driven trapped electron modes (TEM).
These are characterized by their respecitve critical gradient thresholds above which turbulent fluxes rapidly increase.
In tokamaks, it is commonly found that while ETG and ITG are stabilized by an increasing density gradient, TEM activity strongly increases and effectively prevents density profiles from steepening further \cite{Ernst04, Ernst16}.
This effect is commonly called profile consistency, resilience or stiffness, and has been reproduced by nonlinear kinetic turbulence simulations.

The magnetic geometry of the Wendelstein 7-X (W7-X) stellarator has been tailored for low neoclassical (NC) transport \cite{Grieger92, Dinklage18}, which allows transport driven by microinstabilities to play a significant role in the regulation of particle and energy fluxes across the confining magnetic field.
This paper reports on core density turbulence investigations in the W7-X stellarator and demonstrates that in peaked density profile scenarios, ion-scale turbulence and its associated transport is suppressed, supporting improved plasma confinement.

\paragraph{Turbulence evolution in peaked density scenarios}

\begin{figure}
	\includegraphics[width=\hsize]{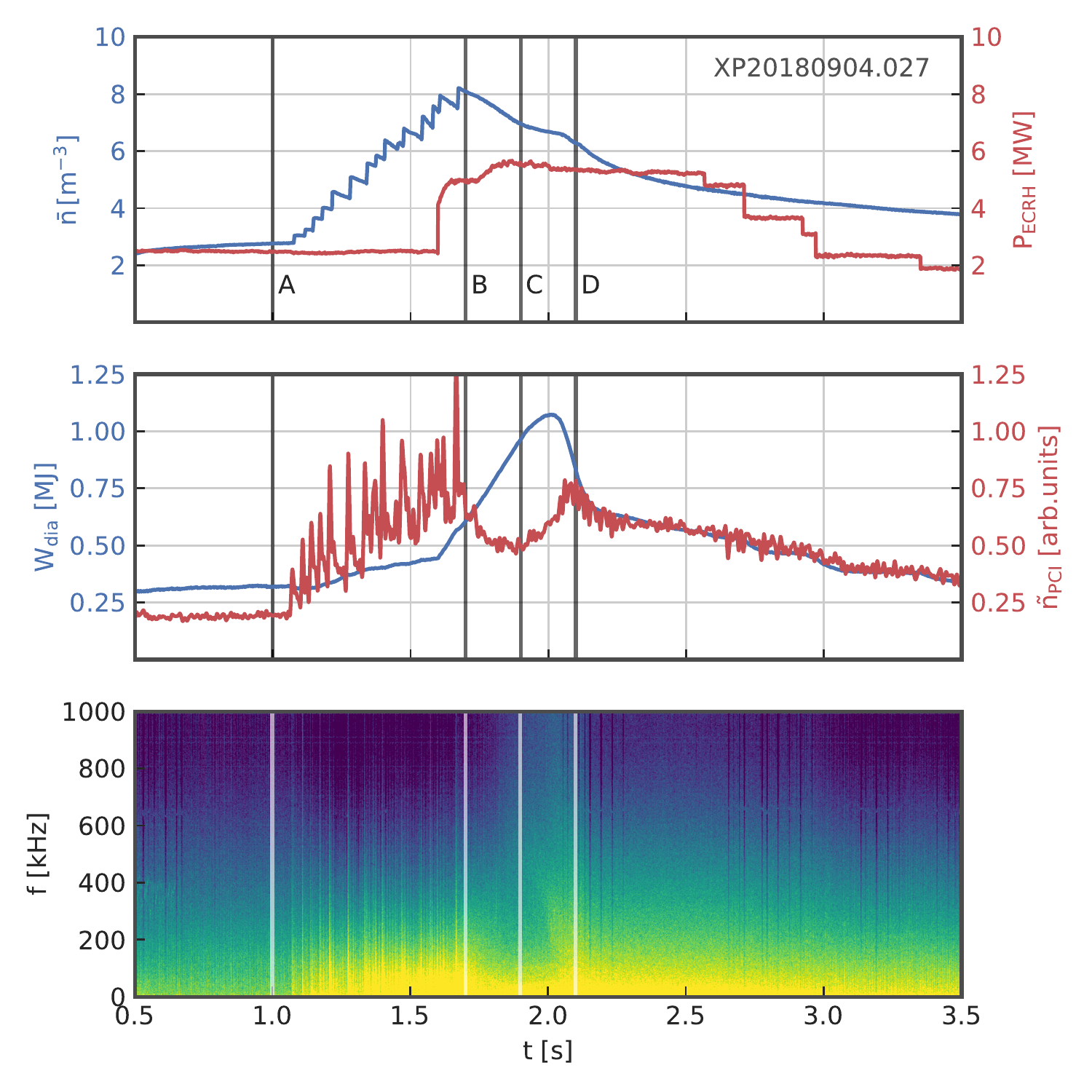}
	\caption{
		W7-X experiment overview showing the line-integrated density $\bar{n}$ and heating power $P_\mathrm{ECRH}$,
		the line-integrated density fluctuations $\tilde{n}$ measured by PCI and diamagnetic energy $W_\mathrm{dia}$, and
		a spectrogram of PCI fluctuations.
		Time points A-D indicated by vertical lines.
	}
	\label{fig:dis_overview}
\end{figure}

\begin{figure}
	\includegraphics[width=\hsize]{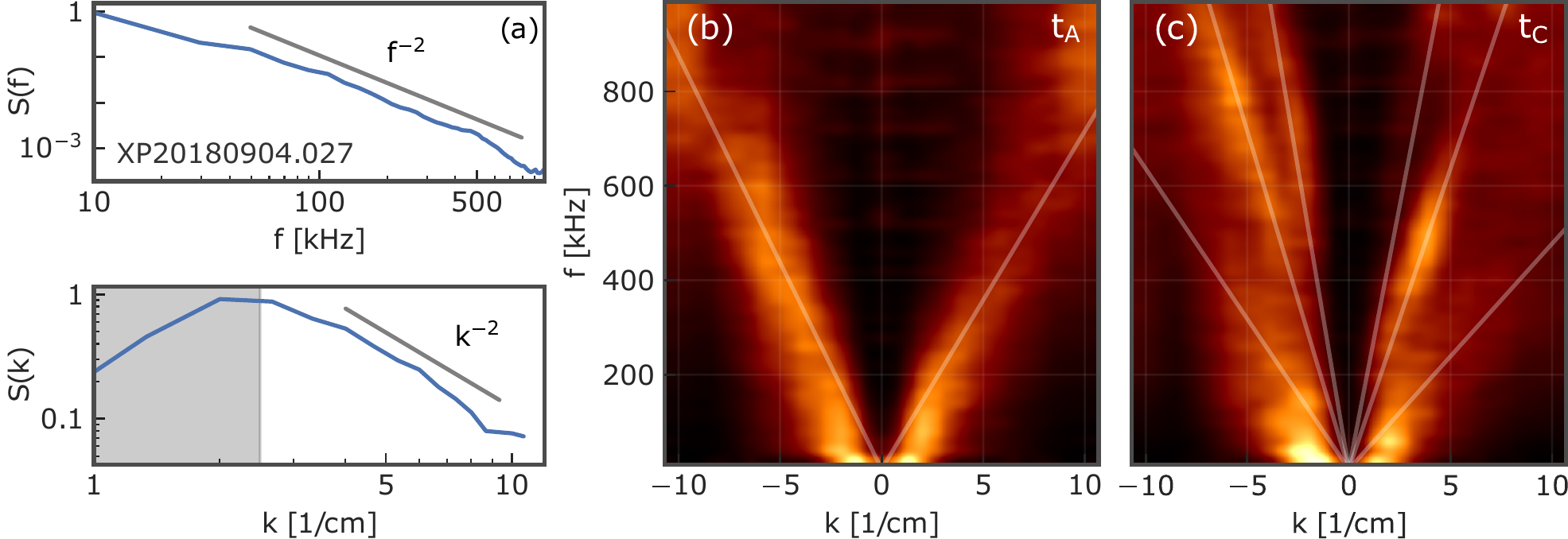}
	\caption{
		(a) Frequency and wavenumber spectra of PCI density fluctuation power at $t_\mathrm{A}$,
		(b) and (c) Frequency-wavenumber spectra at $t_\mathrm{A}$ and $t_\mathrm{C}$, respectively. White lines indicate dominant phase velocities.}
	\label{fig:kf}
\end{figure}

Plasma density peaking in W7-X is influenced both by cryogenic pellet injection \cite{Baldzuhn19} or neutral beam injection (NBI) \cite{Ford20}.
In both discharge scenarios discussed here, initial plasma breakdown occurs by electron cyclotron heating (ECH) only.
For the pellet-fueled scenario (Fig.~\ref{fig:dis_overview}) a sequence of pellets is injected into an initial line-averaged density of $\bar{n}=3\cdot 10^{19}\,\mathrm{m}^{-3}$ between $t_\mathrm{A}=1.0\,\mathrm{s}$ and $t_\mathrm{B}=1.7\,\mathrm{s}$, accompanied by an increase of the heating power.
The plasma density increases and reaches its peak value at $t_\mathrm{B}$ with $\bar{n}=8\cdot 10^{19}\,\mathrm{m}^{-3}$, followed by a smooth decrease.
During this period, and distinctly after the end of the pellet sequence, the plasma stored energy increases sharply from $W_{\mathrm{dia}} = 0.3\,$MJ to a peak exceeding 1\,MJ, a value that has not been achieved in gas-fueled W7-X discharges at equal densities \cite{Bozhenkov20}, thereby indicating strongly improved plasma confinement which lasts until $t_\mathrm{D}$ \cite{Baldzuhn20}.

The primary diagnostic for the detection of core turbulent density fluctuations is phase contrast imaging (PCI) \cite{Edlund18}.
PCI provides a signal whose magnitude is proportional to the absolute, line-integrated density fluctuations along a line of sight from the inboard to outboard side through the magnetic axis, i.e. $I\propto \int \tilde{n} dl$ \cite{Porkolab06}.
By imaging 32 poloidally separated lines of sight, fluctuations are resolved in both frequency and perpendicular wavenumber space $k_\perp$ on scales relevant to ion-scale ($k_\perp\rho_s\approx1$) turbulence.
Fig.~\ref{fig:dis_overview} shows the frequency-integrated PCI fluctuation amplitude $\tilde{n}_{\mathrm{PCI}}$ as well as its spectrogram from a single line of sight.
Before and long after pellet injection ($t<t_\mathrm{A}$ and $t>t_\mathrm{D}$), the turbulent fluctuation spectra are broadband and featureless, and their amplitude generally scales with $W_\mathrm{dia}$.
The pellet injection process itself creates large density perturbations that show up as distinct spikes.
Between $t_\mathrm{B}$ and $t_\mathrm{D}$, fluctuations decrease signficantly while the stored energy increases faster than the density decreases.
$\tilde{n}_{\mathrm{PCI}}$ reaches a minimum at $t_\mathrm{C}$, after which broadband fluctuations set in that rapidly descend in frequency spread as they increase in magnitude.
As fluctuations increase, $W_\mathrm{dia}$ eventually decreases, and fluctuation levels and spectra finally return to values in line with results from gas-fueled discharges.

\begin{figure}
	\includegraphics[width=0.49\hsize]{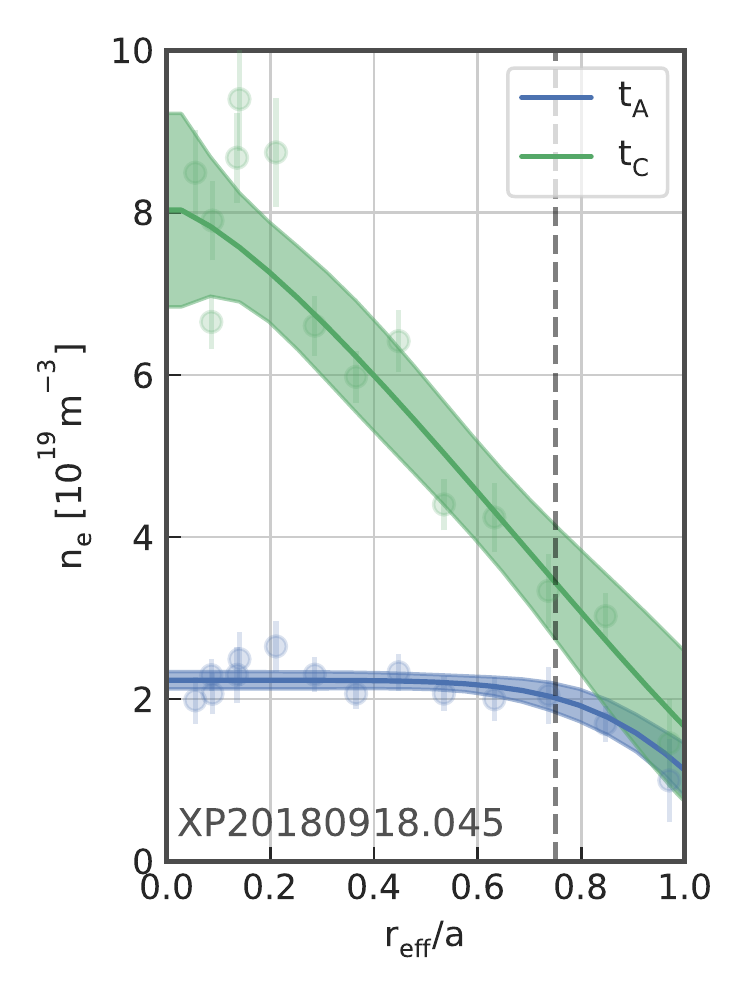}
	\includegraphics[width=0.49\hsize]{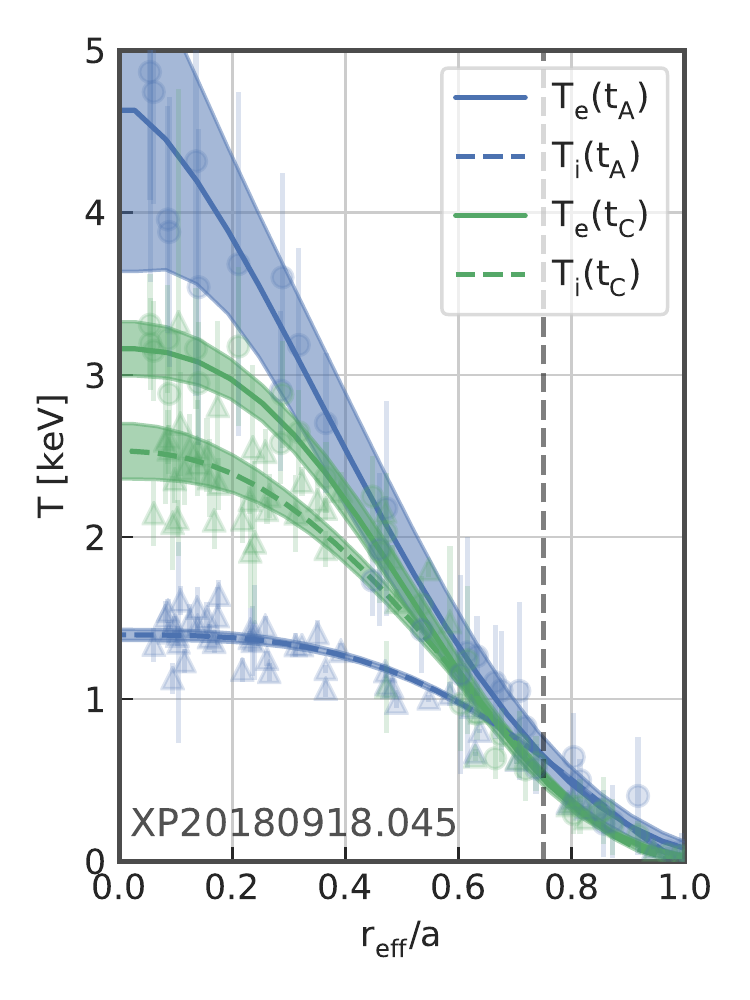}
	\caption{Radial profiles of electron density, ion and electron temperatures before and after density peaking.}
	\label{fig:profiles}
\end{figure}

Figure \ref{fig:kf} (a) displays the fluctuation amplitude spectrum over poloidal wavenumbers $k_\perp$ and frequencies $f$ at the time point $t_\mathrm{A}$, showing broadband power-law behavior (when omitting the shaded low-$k$ instrument limit at $k_\perp<2.5\,\mathrm{cm}^{-1}$).
The fluctuation power is predominantly located at ion scales, with $k_\perp \rho_s = 1$ corresponding to $k_\perp \approx 4\,\mathrm{cm}^{-1}$. 
Figures \ref{fig:kf} (b) and (c) show the combined frequency-wavenumber spectra, normalized to unity at each frequency.
Power is centered around a constant phase velocity $v_{\mathrm{ph}}=2\pi f/k_\perp$ in both positive and negative direction as denoted by white lines, where the sign of $k$ corresponds to the propagation direction in diagnostic coordinates.
We associate the phase velocities in Fig.~\ref{fig:kf} (b) with fluctuations propagating with the $E\times B$ drift, where the two branches correspond to the inboard and outboard side.
Both sides are regions of bad curvature at the toroidal location of the PCI in W7-X in which ion-scale turbulence can develop and are picked up along the single line of sight.
The $k$-$f$ spectra in Fig.~\ref{fig:kf} (c) at $t_\mathrm{C}$, when confinement is improved, show reproducible behavior not observed during regular gas-fueled discharges: several distinct and simultaneous phase velocities indicate a qualitative change in turbulence, which persists from $t_\mathrm{B}$ to shortly before $t_\mathrm{D}$, when a single dominant phase velocity is recovered and the situation is again analogous to that before pellet injection.

The period of improved confinement and reduced fluctuations is accompanied by a significant steepening of the density profiles in the plasma core as documented in Ref. \cite{Bozhenkov20}.
Figure \ref{fig:profiles} shows radial density and temperature profiles of an experimental program phenomenologically similar to that shown in Fig.~\ref{fig:dis_overview}, before pellet injection ($t_\mathrm{A}$) and when density fluctuations are at a minimum ($t_\mathrm{C}$).
The density profiles during gas-fueled ECH discharges are relatively flat throughout the plasma core, with a normalized gradient of $a/L_n=-(a/n) \, dn/dr=0.41\pm0.36$ at the chosen evaluation radius of $r_\mathrm{eff}/a=0.75$ (dashed lines), where $a$ is the minor radius.
Pellet injection steepens this gradient to $a/L_n=1.13\pm0.49$ and extends it deep into the core, and this peaking persists throughout the period of improved confinement.
The associated electron temperature profiles are peaked with a reduced central value after pellet injection following the large increase in density.
Outside of $r\mathrm{_{eff}}/a=0.75$, $T_\mathrm{i}\approx T_\mathrm{e}$, but there is a departure of the ion from the electron temperature further in the core.
During gas-fueled discharges, the ion temperature is essentially flat inside the half radius, and only marginally varies in profile shape and central value despite large variations in density and electron temperature \cite{Bozhenkov20}.
During improved confinement, $T_\mathrm{i}$ rises to match $T_\mathrm{e}$ further into the core, steepening its gradient and reaching the highest central values observed in the experiment.

\paragraph{Transport assessment}

\begin{figure}
	\includegraphics[width=0.49\hsize]{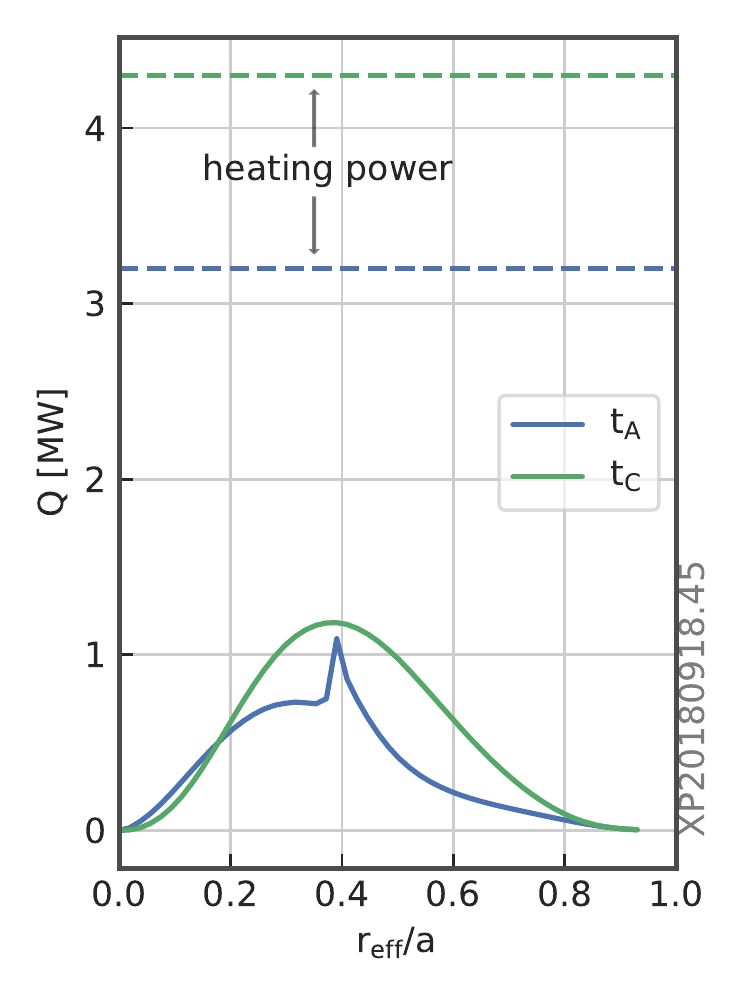}
	\includegraphics[width=0.49\hsize]{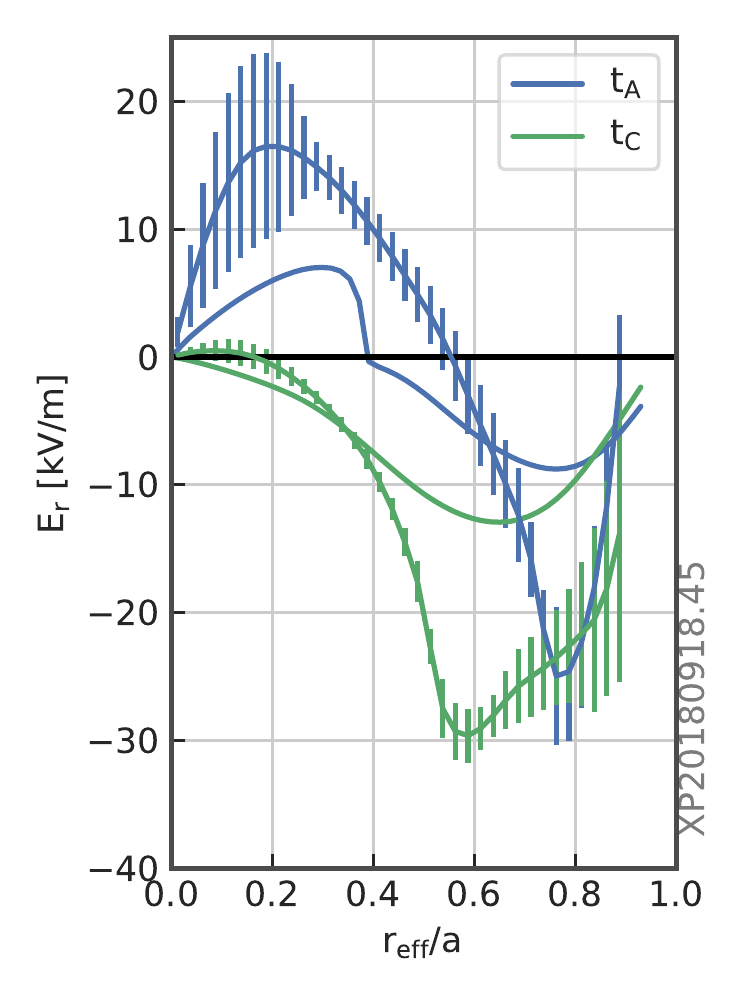}
	\caption{Total neoclassical heat fluxes for time points $t_\mathrm{A}$ and $t_\mathrm{C}$ as shown in Fig.~\ref{fig:profiles} and associated radial electric field profiles from neoclassical modeling (solid) and measurements (error bars).}
	\label{fig:NCtransport}
\end{figure}

The NC radial fluxes are calculated by solving the drift-kinetic equation using the DKES \cite{Hirshman86, Vanrij89} code.
At the high densities that W7-X aims to achieve, interspecies collisions are expected to make the ion and electron temperatures almost equal. In such plasmas an inward radial electric field $E_\mathrm{r}$ arises, the so-called ion root \cite{Galeev79, Pablant20}.
Another situation arises when the density is not high enough for collisions to efficiently couple $T_\mathrm{i}$ to the high $T_\mathrm{e}$ produced by the electron heating in the experiment.
In this case, an outward (electron root) radial electric field can result, which is often seen to occur in the central portion of the plasma \cite{Yokoyama07, Dinklage18}.

Figure \ref{fig:NCtransport} shows the NC heat fluxes $Q$ and ambipolar electric fields $E_r$ calculated for the experimental profiles shown in Fig.~\ref{fig:profiles}.
At $t_\mathrm{A}$ the central plasma is in electron root, because $T_\mathrm{e}$ is much larger than $T_\mathrm{i}$, while at $t_\mathrm{C}$ the entire cross section is in ion root with a significantly larger electric field magnitude towards the edge.
This behavior is qualitatively confirmed by measurements of poloidal flows, where the deduced $E_r$ show a transition from electron to ion root and a strong magnitude increase in the temperature gradient region (error bars in Fig.~\ref{fig:NCtransport}(b) only reflect photon statistics, systematic uncertainties are significantly larger \cite{Pablant20}).
A comparison of the NC energy fluxes integrated over the flux surface area (solid lines) with the input heating power (dashed lines) yields that NC transport can only account for at most 30\% of the total heat flux at both time points (where $dW/dt\approx0$, radiation is low and comparable between both cases and charge exchange losses are unknown).
The observed improved confinement after pellet injection can therefore not stem from a change of NC transport, but is attributed to a considerable reduction of anomalous heat flux.

\begin{figure}
	\includegraphics[width=\hsize]{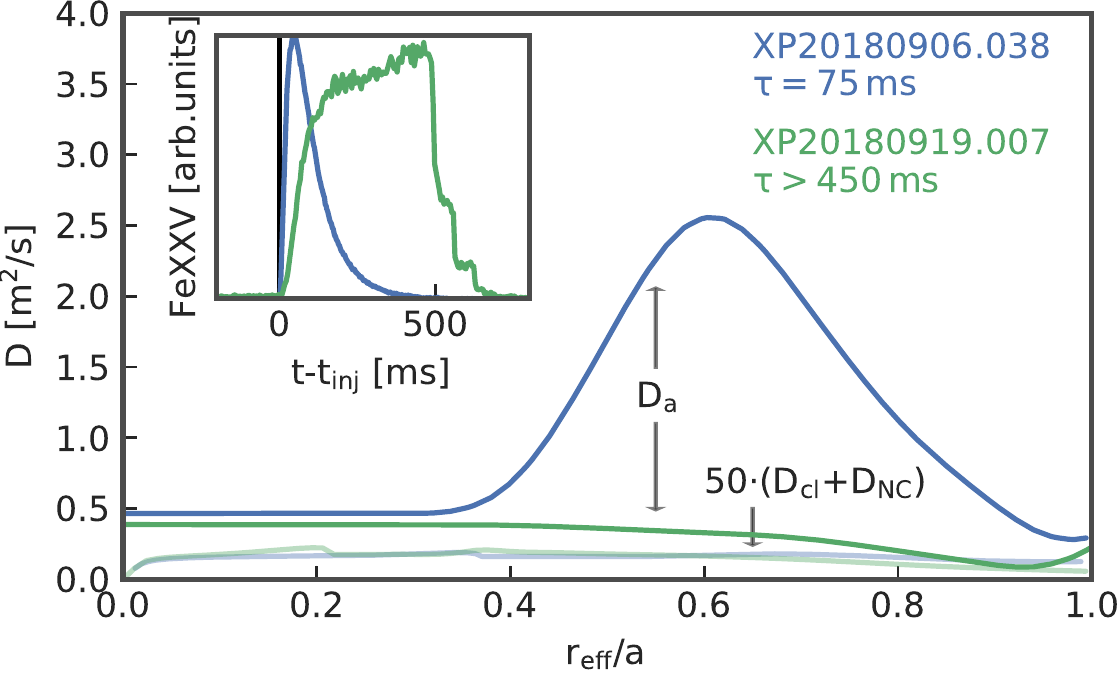}
	\caption{Impurity time traces (inset) and resulting classical and neoclassical (faint) and anomalous (solid) impurity diffusion coefficient profiles in two experiments.}
	\label{fig:strahl}
\end{figure}

A similar finding is obtained when the particle transport, as measured by the confinement time of impurities injected via the laser blow-off system \cite{Wegner2018}, is considered.
Figure \ref{fig:strahl} shows the measured confinement times of iron impurities injected during two experiments, represented by the inset FeXXV time traces, as well as the respective diffusion coefficients derived from STRAHL calculations \cite{Behringer87}.
Situations with increased density gradient have been identified apart from pellet injection at low heating power after step-down of ECH and/or NBI power.
Such a case is shown at 1.2\,MW ECH power 0.5\,s after NBI shutdown (green), and is contrasted with a regular ECH discharge (3\,MW) with flat density profiles (blue).
As with the heat flux, it is generally observed that the sum of the classical and neoclassical components (faint lines) only account for a small fraction of the total radial impurity flux. The anomalous component, especially the diffusion (solid), must be much larger in order to explain the short confinement times on the order of 100\,ms, indicative that turbulence also contributes significantly to radial impurity transport \cite{Geiger19}.
In contrast, anomalous impurity particle diffusion decreases considerably in the peaked density discharge, resulting in an impurity confinement time longer than the remainder of the discharge 500\,ms after injection.
As in the pellet case, accompanying PCI measurements exhibit the same characteristic signatures of a reduced fluctuation amplitude and multiple phase velocities, which suggests that these processes are linked.
In conclusion, we find that the NC transport assessment and particle diffusion measurements both show that anomalous transport dominates radial heat and particle flux, and that a centrally peaked density profile reduces its severity.

\paragraph{Gyrokinetic simulations}

\begin{figure}
\includegraphics[width=0.42\hsize]{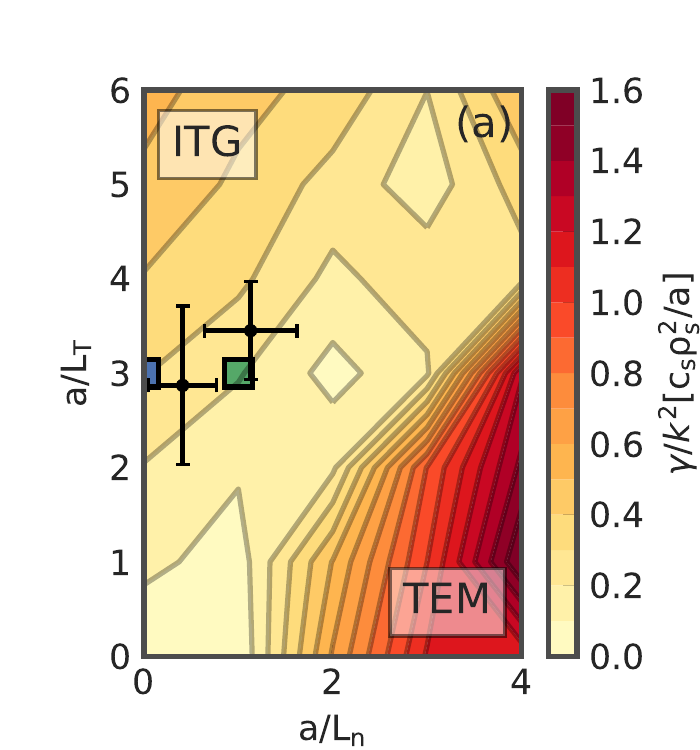}
\includegraphics[width=0.56\hsize]{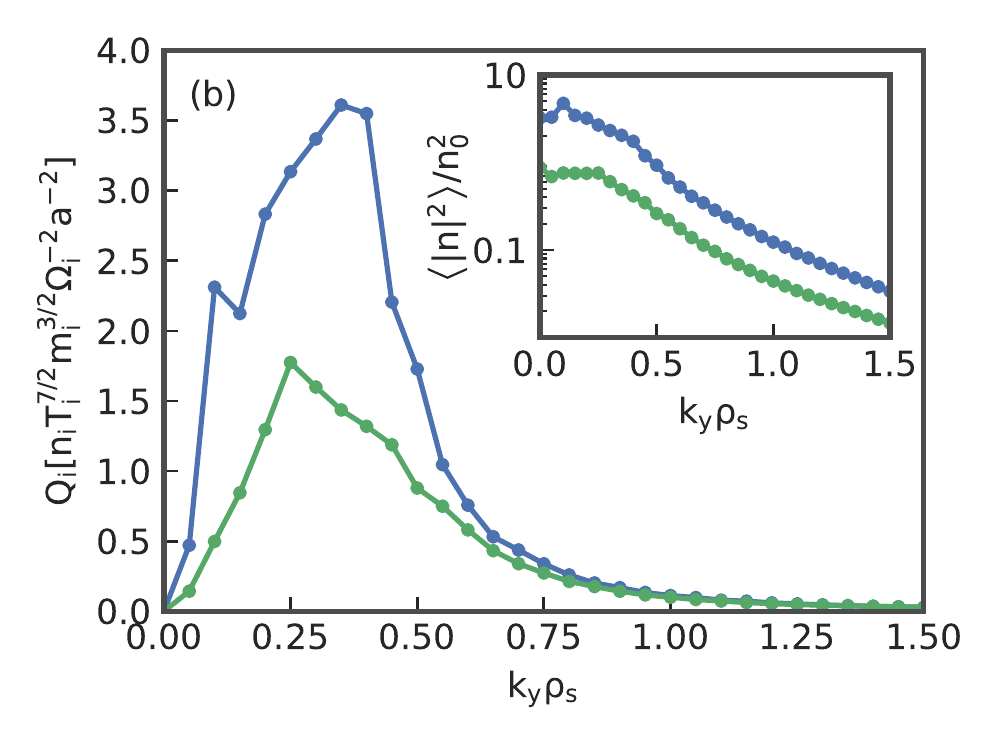}
\caption{
(a) Map of maximum effective diffusivity $\gamma/k^2$ for various gradients in the ``standard'' magnetic configuration of W7-X with experimental points before and after pellet injection (error bars) and values used for nonlinear simulations (squares).
(b) Gyro-Bohm normalized nonlinear ion heat flux and density fluctuation spectrum for those two sets of gradients.
}
\label{fig:simulations}
\end{figure}

These experimental findings are compared to results from the flux tube version of the simulation code \texttt{GENE}~\cite{Jenko00} which numerically solves the coupled nonlinear system of gyrokinetic equations in toroidal magnetic configurations featuring nested magnetic flux surfaces.
Since the three-dimensional magnetic field geometry of W7-X has been shown to prevent significant transport from electron-scale turbulence when ion-scale turbulence is present \cite{Plunk19}, gyrokinetic simulations are done which cover the two major ion-scale instabilities, namely the ion temperature gradient (ITG) and trapped electron mode (TEM), and $a/L_{T_\mathrm{e}}=0$ is assumed.
Results for ITG and TEM are compiled in Fig.~\ref{fig:simulations} (a) for the ``standard'' magnetic configuration at $r\mathrm{_{eff}}/a=0.6$ in Fig.~\ref{fig:profiles}.
Here, the normalized density and ion temperature gradients $a/L_n$ and $a/L_{T_\mathrm{i}}$ are varied, and $T_\mathrm{i}=T_\mathrm{e}$ as well as $a/L_{T_\mathrm{e}}=0$ are assumed, while also neglecting collisions and electromagnetic effects\cite{Alcuson20}.
Invoking the mixing length estimate for a rough prediction of effective diffusivity, the plots show the maximum of $\gamma/k^2$ ($\gamma$ is the linear growth rate).
A ``stability valley'' of low effective diffusivity is seen in the intermediate region of $\eta_i=L_{n}/L_{T_\mathrm{i}} \approx1$ \cite{Helander15}, which has been shown to result from the combination of two beneficial effects:
A reduction in $\eta_i$ is well-known to result in reduced ITG drive, though an increased density gradient in tokamaks leads to strong TEM growth which counterbalances this reduction.
It has however been shown \cite{Proll12, Helander13} that W7-X belongs to a class of magnetic configurations (``maximum-$J$'') that are resilient to these modes, and large TEM growth rates are therefore only observed at very high density gradients.
Inserting the measured normalized gradient values from the profiles shown in Fig.~\ref{fig:profiles} at $r_\mathrm{eff}/a=0.75$ into Fig.~\ref{fig:simulations} (a), we see that the quasi-linear estimate predicts a reduction in turbulent heat flux with increasing density gradient.

This is further supported by nonlinear simulations for two representative parameter sets before and after density gradient steepening.
These are performed on the $r/a=0.75$ surface along the flux tube crossing the outer bean shaped plane of the high-mirror (HM) configuration of W7-X, and the resolution is set to $(n_z,n_{kx},n_{ky},n_{v},n_w,k_{y,min})=(256, 64, 96, 48, 10, 0.05)$, where $nz$ is the number of points along a field line, $n_{kx}$ and $n_{ky}$ are the number of radial and perpendicular Fourier modes, $n_v$ and $n_w$ are the number of points in parallel and magnetic moment aligned velocity space, and $k_{y,min}$ is the smallest resolved normalized wavenumber.
Electrons are treated kinetically in order to allow for growth of TEM modes.
The gradients are chosen as $a/L_{T_\mathrm{i}}=3$ and $a/L_n=0$ or 1 (squares in Fig.~\ref{fig:simulations}(a)) with $T_\mathrm{e}=T_\mathrm{i}$, while the electron temperature gradient is set to zero in both cases in order to avoid convergence issues due to electron-scale turbulence.
Fig.~\ref{fig:simulations}(b) shows the resulting nonlinear gyro-Bohm normalized ion heat flux spectrum (which dominates the total flux).
It is reduced at all scales, corresponding to a drop of the wavenumber-integrated flux from $Q_i/Q_{GB}=30$ to 15 by the introduction of a density gradient.
Additionally, the inset shows that density fluctuations also decrease significantly throughout the scales detectable by the PCI diagnostic, in line with the experimental observations described above.
Though the nonlinear simulations were performed in the HM configuration, the same effect is observed in HM experiments, and in simulations using the standard configuration (albeit at different gradients, see \cite{Xanthopoulos20}).
Both configurations exhibit the maximum-J feature (to different degrees), which is key for the reduction of TEM transport in the case of central density peaking.
It is worth noting that simulations performed with equal gradients in tokamak geometry which does not benefit from the stabilizing properties of maximum-$J$ configurations instead show a strong increase in the total nonlinear heat flux \cite{Helander15}.

In summary, we find that discharges in W7-X with peaked plasma density profiles are accompanied by an overall reduction of both the turbulence amplitude and anomalous impurity diffusion as measured by two separate diagnostic systems.
The consistent presence of multiple modes in the PCI spectra may imply a mechanism for reducing turbulent transport analogous to the I-mode in tokamaks \cite{Hubbard17,Ryter16} in the sense that density fluctuation modes are observed during a change of confinement regimes.
Neoclassical transport calculations show that heat flux is governed largely by turbulent transport, while gyrokinetic simulations suggest that a reduction of turbulent heat flux is due to the unique maximum-$J$ property of W7-X which limits the severity of TEMs while ITG modes are stabilized \cite{Proll12}.
We note that other stabilization mechanisms governed by the large observed increase of the radial electric field magnitude as well as its shear have been shown to be effective in stellarators \cite{Hirsch08}, and the impact of its magnitude through $E \times B$ fluctuation advection in W7-X is investigated separately \cite{Xanthopoulos20}.
An estimate of the shearing rate from the measured $E_r$ shows that it can be comparable to the growth rates at $r\mathrm{_{eff}}/a>0.6$ and may therefore further impact transport \cite{Kinsey05}.
These results demonstrate that the optimized magnetic geometry of W7-X is favorable to reach improved confinement scenarios by tailoring the plasma density profile and thereby reducing turbulence.

\paragraph{Acknowledgements}
The \texttt{GENE} simulations have been conducted on supercomputers Marconi (Italy), Helios of IFERC-CSC (Japan) and MPCDFs Hydra and Cobra (Garching, Germany).
This work is partly sponsored by the US Department of Energy, Office of Fusion Energy Sciences under grant number DE-SC0014229.
This work has been carried out within the framework of the EUROfusion Consortium and has received funding from the Euratom research and training programme 2014-2018 and 2019-2020 under grant agreement No 633053. The views and opinions expressed herein do not necessarily reflect those of the European Commission.

\bibliographystyle{apsrev4-1}
\bibliography{references}

\end{document}